\documentclass[10pt,twocolumn,preprintnumbers,amsmath,amssymb,nofootinbib
,superscriptaddress, floatfix]{revtex4-2}
\usepackage{graphicx,longtable,mathrsfs,color,array}
\usepackage[hidelinks]{hyperref}
\usepackage[usenames,dvipsnames]{xcolor} 
\usepackage{amssymb,amsmath,mathtools,mathrsfs,slashed} 
\usepackage{epsfig,subfigure,placeins,float} 
\usepackage{booktabs,longtable,ctable,multirow} 
\usepackage{exscale,relsize} 
\usepackage[normalem]{ulem} 
\usepackage{enumerate}
\usepackage{times, mathptmx} 
\usepackage[utf8]{inputenc}
\usepackage{color}
\usepackage{graphicx}
\usepackage{graphicx,graphics}
\graphicspath{{Figures/}{../Figures/}}
\usepackage{bm}
\usepackage{tabularx}
\usepackage{rotating}
\usepackage{units}
\allowdisplaybreaks[1]
\usepackage{hhline}
\usepackage{soul}
\usepackage{xspace}

\newcommand{\lcdm}{$\Lambda$CDM\xspace}
\newcommand{\wcdm}{$w$CDM\xspace}
\newcommand{\mnras}{Monthly Notices of the RAS}
\newcommand{\jcap}{Journal of Cosmology and Astroparticle Physics}
\newcommand{\apjl}{Astrophysical Journal, Letters}
\newcommand{\apjs}{Astrophysical Journal, Supplement}
\newcommand{\aj}{Astronomical Journal}

\newcommand{\araa}{Annual Review of Astron and Astrophysics}
\newcommand{\aap}{Astronomy and Astrophysics}
\newcommand{\pasj}{Publications of the ASJ}

\begin{document}

\title{The impact of the Universe's expansion rate on constraints on modified growth of  structure}

\author{Jaime Ruiz-Zapatero}
\email{jaime.ruiz-zapatero@physics.ox.ac.uk}
\affiliation{Astrophysics, University of Oxford, DWB, Keble Road, Oxford OX1 3RH, UK}
\author{David Alonso}
\affiliation{Astrophysics, University of Oxford, DWB, Keble Road, Oxford OX1 3RH, UK}
\author{Pedro G. Ferreira}
\affiliation{Astrophysics, University of Oxford, DWB, Keble Road, Oxford OX1 3RH, UK}
\author{Carlos Garcia-Garcia}
\affiliation{Astrophysics, University of Oxford, DWB, Keble Road, Oxford OX1 3RH, UK}

\begin{abstract}
In the context of modified gravity, at linear level, the growth of structure in the Universe will be affected by modifications to the Poisson equation and by the background expansion rate of the Universe. It has been shown that these two effects lead to a degeneracy which must be properly accounted for if one is to place reliable constraints on new forces on large scales or, equivalently, modifications to General Relativity. In this paper we show that current constraints are such that assumptions about the background expansion have little impact on constraints on modifications to gravity. We do so by considering the background of a $\Lambda$ Cold Dark Matter ($\Lambda$CDM) universe, a universe with a more general equation of state for the dark energy, and finally, a general, model-independent, expansion rate.  We use Gaussian Processes to model modifications to Poisson's equation and, in the case of a general expansion rate, to model the redshift dependent Hubble rate. We identify a degeneracy between modifications to Poisson's equation and the background matter density, $\Omega_M$, which can only be broken by assuming a model-dependent expansion rate. We show that, with current data, the constraints on modifications to the Poisson equation via measurements of the growth rate range between $10-20\%$ depending on the strength of our assumptions on the Universe's expansion rate.

\end{abstract}

\maketitle


\section{Introduction}

The growth of structure in the Universe is a sensitive probe of fundamental physics \cite{Ferreira:2010sz,Ferreira:2019xrr}. It is  driven by  gravitational collapse but is also sensitive to additional forces which may be undetectable on smaller, laboratory scales. It has been shown that measurements of the {\it rate} of growth of structure can be used to test gravity and constrain, as yet, elusive fifth forces \cite{Adelberger:2003zx}.

To be specific, the motion of matter in the universe can, in general, be subjected to an effective force, ${\vec F}_{\rm eff}$ of the form
\begin{eqnarray}
{\vec F}_{\rm eff}=-{\vec\nabla}\Psi_N-{\vec\nabla}\Psi_5.
\end{eqnarray}
Here, $\Psi_N$ is the Newtonian potential and $\Psi_5$ is the potential for a possible long range force that co-exists with gravity on large scales. The properties of $\Psi_5$ may depend on the state of the Universe (its expansion rate, the fractional energy densities of its different constituents) or even on local environmental properties~\cite{Clifton:2011jh,Joyce:2014kja}. Thus $\Psi_5/\Psi_N$ will, generally, be a function of space and time.

If we restrict ourselves to purely long range forces with no environmental dependence, we can define a generalized Newtonian potential, $\Psi\equiv\Psi_5+\Psi_N$. In an expanding Universe with scale factor, $a$,  $\Psi$ satisfies a Newton-Poisson equation on sub-horizon scales
\begin{eqnarray}
\nabla^2\Psi=4\pi G \mu a^2 {\bar \rho}\delta \, ,
\end{eqnarray}
where $G$ is Newton's constant, ${\bar \rho}$ is the background energy density of non-relativistic matter, $\delta$ is the density contrast and $\mu$ is a function of time only.  The relative amplitude of the new force, at any moment in time, is given by $\mu-1$.

From the linearized Newton-Poisson, continuity and Euler equations one can derive and evolution equation for the growth rate of structure, $f\equiv d\ln \delta/d\ln a$, given by
\begin{eqnarray}
f'+f^2+\left(1+\frac{d\ln aH}{d\ln a}\right)f=\frac{3}{2}\mu\Omega_M(a) \, , \label{eq:growth}
\end{eqnarray}
where prime is derivative with regards to $\ln a$, $H$ is the Hubble rate and $\Omega_M(z)$ is the fractional energy density in matter as a function of redshift \cite{1980lssu.book.....P,Ferreira:2010sz,Baker14}. 
Thus, as we can see, the evolution of $f$ depends on $\mu$. This means that, in theory, one can use measurements of the growth rate to constrain the presence of fifth forces.

The situation is, of course, more complex. The evolution of the growth rate depends on the evolution of $H$ and $\Omega_M(a)$. 
The latter quantity depends, through the Einstein field equations, on $H(a)$ so that
\begin{eqnarray}
\Omega_M(a)=\frac {\Omega_M(0)H^2_0}{a^3H^2}  \, .
\end{eqnarray}
Thus, measurements of the growth rate can be used to place constraints on the time evolution of $\mu$ and $H$, and on the fractional matter density today, $\Omega_M(0)$ (for ease of notation, we will now refer to it as $\Omega_M$ with no argument). But this means that constraints on these various quantities are intertwined and, unless we have independent methods for pinning down $H$ and $\Omega_M$, they will hamper our ability to determine $\mu$. 

This degeneracy between $\mu$ and the expansion history (encapsulated in $H$, for example) was discussed in \citet{Simpson:2009zj}. There, it was shown that there is a degeneracy between $ \gamma  \equiv \partial \ln f / \partial \ln \Omega_M $ and the equation of state of the dark energy component, $w\equiv P_{\rm DE}/\rho_{\rm DE}$, where $\rho_{\rm DE}$ ($P_{\rm DE}$) is the energy density (pressure) of the substance responsible for the accelerated expansion of the Universe at late time (the dark energy).  In \citet{Baker14}, explicit expressions for the degeneracy between $\mu$ and $w$ were found using the linear response approach. 

Most attempts at constraining $\mu(z)$ have assumed a Universe in which the accelerated expansion at late time is driven by a cosmological constant: the $\Lambda$ Cold Dark Matter ($\Lambda$CDM) model \cite{Planck,Joudaki:2017zdt,Mueller:2016kpu}. A further assumption is that $\mu(z)$ can be modelled in terms of a simple function with one (or at most two) parameters \cite{Planck}. In a few cases, a more general form for $\mu(z)$ has been assumed with a few independent values at different redshifts (for a notable example see \citet{Joudaki:2017zdt}). Alternatively, model specific time dependences for $\mu(z)$ have been assumed arising from theoretical arguments, either from the Effective Field Theory of dark energy \cite{Espejo:2018hxa} or from choices for the underlying model of gravity (such as shift symmetric scalar tensor gravity and its extensions \cite{Traykova:2021hbr}). Most of these attempts at constraining $\mu(z)$ have side-stepped the issue of the degeneracy described above although we highlight \citet{Raveri:2021dbu} in attempting to obtain model-independent constraints.

In this paper we will explore how current constraints on $\mu$ are affected by our assumptions about the expansion rate of the Universe. In particular, we will see how more or less restrictive assumptions about the parametric form of $H(z)$ impact the uncertainty with which we can determine $\mu(z)$. In the limit in which we do not assume a parametrized form for $H(z)$ we will show that a fundamental degeneracy between $\Omega_M$ and $\mu(z)$ manifests itself and, in that regime, we must resign ourselves to constraining the combination $\Omega_M\mu(z)$.

The structure of this paper is as follows. In Section~\ref{Sect: a Gaussian process for muz} we present the main method of this paper, the use of a Gaussian process as a model-independent parametrisation of $\mu(z)$. In Section~\ref{Sect: Interpreting a Gaussian Process} we discuss how to interpret said Gaussian process. In Section~\ref{Sect: Observables and data sets} we describe the cosmological observables and the associated data sets which we will use to find the constraints in this paper. In Section~\ref{Sect: model-dependent} we present our constraints on $\mu(z)$ and how they depend on what we assume as a model for the background evolution; we will focus on $\Lambda$CDM and its extension $w$CDM, in which we assume an (possibly time varying) arbitrary equation of state, $w$. In Section~\ref{Sect: Double GP} we completely free the background evolution and model $H(z)$ as a Gaussian Process. This gives rise to a strong degeneracy between $\Omega_M$ and $\mu(z)$ and we can only constraint ${\tilde \mu}(z)=\Omega_M\mu(z)$. In Section~\ref{Sect: Conclusions} we discuss both our finding about the role of Gaussian processes in cosmological analysis and the constraints we have found on $\mu(z)$.

\section{A Gaussian Process for \texorpdfstring{$\mu(z)$}{Lg}} \label{Sect: a Gaussian process for muz}

The goal of this work is to quantify the uncertainty in our knowledge of $\mu(z)$. The quality of this constraint will depend on both the quality of the data and the assumptions we make about the underlying cosmology through the expansion rate. We want to assume that we have no prior knowledge of the time dependence of $\mu(z)$, apart from the fact that it is relatively smooth. Thus, we choose to model $\mu(z)$ as a Gaussian Process (GP).

GP's have been extensively used in astrophysics as tools to model different quantities in an agnostic way \citep{1802.01505, 1907.10813, Liao_H_gp, Bonilla_H0_gp, S8_gp, Huillier18, 2011.11645, 2007.05714, Shafieloo18, 1902.0942, 1806.02981, 2105.01613, Ruiz_Zapatero_21}. Fundamentally, a GP is a collection of random variables (nodes), each of them sampled from a multivariate Gaussian distribution with a non-diagonal covariance \citep{1204.2832}. Thus a GP; $g(\textbf{x})$ where $\textbf{x}$ is a arbitrary vector representing the position of the nodes,  is fully specified by a mean function; $m(\textbf{x}) \equiv \mathcal{E}[g(\textbf{x})]$ - where $\mathcal{E}[\cdots]$ is the expectation value over the ensemble -  and a covariance function; $k(\textbf{x}, \textbf{x'}) \equiv \mathcal{E}[(g(\textbf{x})-m(\textbf{x}))(g(\textbf{x'})-m(\textbf{x'}))]$. In combination, the mean and covariance functions determine the statistical properties of the random variables that define the family of shapes that the GP can take. In our case, we chose $\mu(a) = 1$ as the mean of our GP since this is the value corresponding to GR. For the covariance function, we choose a square exponential covariance function, defined as
\begin{eqnarray} 
k\left[g(x),g(x')\right]=\eta^2\exp\frac{|x-x'|^2}{2l^2} \, ,  \label{eq:QuadExp}
\end{eqnarray}
where $\eta$ is the amplitude of the oscillations around the mean and $l$ is the correlation length between the GP realizations. This decision was made based on the fact that the square exponential is computationally inexpensive and infinitely differentiable kernel, appropriate to model smooth fluctuations around the mean of the GP.

Given a likelihood $\mathcal{L}(\textbf{y}| \textbf{x}, \boldsymbol{\sigma})$ for a set of data points $\textbf{y}$,  with a set of errors $\boldsymbol{\sigma}$, and a set of random variables $\textbf{x}$, a GP can be employed as a prior over all the possible families of functions used to fit the observations. Observations can then be used to inform the GP posterior (i.e. the statistical properties of the assemble of random variables), $\mathcal{P}(g(\textbf{x})| \textbf{y},\boldsymbol{\sigma})$, which determines the family of functions most consistent with the data. 

Since we do not have direct measurements of $\mu(z)$, we have to infer it from measurements of the growth rate. However, as one can see from Eqn. \ref{eq:growth}, $f\sigma_8$ also depends on $H(z)$ and $\Omega_M$. Thus, we must jointly determine $\mu(z)$, $H(z)$ and $\Omega_M$ in terms of measurements of $f\sigma_8$ and $H(z)$, or derived quantities such as the comoving ($D_M(z)$), luminosity ($D_L(z)$) or the angular diameter ($D_A(z)$) distances, which relate with $H(z)$ via
\begin{eqnarray}  \label{eq:distance}
    (1+z)D_A=\frac{D_L}{1+z}=D_M &=& \int_0^{z}\frac{dz'}{H(z')} \quad.
\end{eqnarray}

In summary, as we can see from Eqns. \ref{eq:growth} and \ref{eq:distance}, computing predictions for our observables will involve a non-linear, non-local mapping between the quantities we are interested in ($\mu(z)$, $\Omega_M$, $H(z)$...) and the data. For example, a measurement of $f\sigma_8$ at a particular redshift, $z$, constrains the {\it history} of $\mu$ up until that redshift and not only the value of $\mu$ {\it at} that redshift.

The fact that the variables of our model are not linearly related to our data has, nonetheless, strong implications. Namely, we will have to sample the GP nodes as individual parameters instead of just constraining the statistical properties of the ensemble. This means that our model will contain of the order of $\mathcal{O}(10^2)$ parameters. This large number of parameters (and hence dimensions) renders traditional parameter space exploration techniques too slow to be feasible. 

In the Metropolis Hastings (MH) sampler \cite{Metropolis, Hastings}, samples are drawn randomly from a proposal distribution. If the new proposal improves the fit to the data it is automatically accepted. If not, the sample has a random chance of being accepted to avoid falling into a local minimum. This means that the chance of the sampler drawing a better sample and thus of the sample being accepted decreases with number of dimensions of the parameter space. This decreasing acceptance-rate of new samples means that the time needed by samplers that randomly explore the parameter space quickly becomes unfeasible as we increase the number of parameters. This is known as the dimensionality curse. 

To remedy this, in this work we make use of the No U-turns sampler (NUTS) \citep{NUTS}, a self-tuning version of the Hamiltonian Monte Carlo (HMC) algorithm  \citep{HMC, Betancourt17}. HMC uses notions of Hamiltonian dynamics to draw trajectories on the parameter space along which the sampler moves. This results in a much greater acceptance rate, and allows HMC to beat the dimensionality curse. Therefore, HMC can efficiently explore parameter spaces with large numbers of dimensions in far less time than MH or nested sampling techniques \citep{NS_Handley}. 

The drawback of HMC is that in order to evolve the Hamiltonian equations of motion it is necessary to compute the derivatives of the likelihood with respect to the parameters. Obtaining such derivatives can be even more expensive than taking additional steps in the chains, especially in high dimensional spaces. In order to overcome this issue, we require an inexpensive way of obtaining derivatives of the likelihood. In this work, we employ the \texttt{Python} package \texttt{PyMC3} \citep{Pymc3} which uses the auto-differentiation \citep{Autodiff18} library \texttt{Theano} \citep{Theano} to obtain the gradient of our model with respect to our parameters. This is achieved by drawing a symbolic graph of the model that establishes the relationship between the different variables. 

Finally, our choice of NUTS over the traditional HMC is that in the latter one needs to be able to infer a-priori (or by trial and error) specifications such as for how long the sampler should follow the trajectory or to what precision it needs to be resolved. NUTS can tune these parameters during the burn-in phase of the chains by enforcing that the sampler does not perform a U-turn while following a trajectory, preventing the samples from becoming correlated.

\section{Interpreting a Gaussian Process} \label{Sect: Interpreting a Gaussian Process}

In the previous section we described how to model $\mu(z)$ using a GP. In this section we will discuss how to interpret it. However, as we will see, this is no easy task. The problem fundamentally stems from the fact that a GP is not a single parameter with a singular figure of merit (e.g. the standard deviation), but a vector of parameters. Nonetheless, if we wish to assess how well we can constrain $\mu(z)$ we need to devise a compact and useful way of compressing (and comparing) the information we get from the GP.  

As discussed in Sect. \ref{Sect: a Gaussian process for muz}, the statistical properties of a GP are encapsulated in its mean and covariance matrix. Therefore, if one wishes to measure how constrained a GP is, the first intuition would be to turn to the covariance matrix of the GP's posterior; the multi-dimensional equivalent of the standard deviation. The problem that arises is finding a way to compress such a covariance matrix into a meaningful measurement

A first idea would be to look at the determinant of said covariance matrix. However, the determinant mixes contributions from both the diagonal elements of the matrix; i.e. the standard deviation in each node, and from the off-diagonal elements of the matrix; i.e. the correlations between the nodes, in a non-trivial way that obfuscates its interpretation. One could then think of diagonalizing the covariance matrix. However, since diagonalizing is itself a non-linear operation interpreting the errors of the diagonal basis would be a non-trivial task.

Alternatively, one could take advantage of the so called hyperparameters of the GP. Hyperparameters dictate the values that the nodes are allowed to take and that act as a high level description of the statistical properties of the nodes. The most relevant hyperparameter would be the amplitude of the GP covariance matrix which dictates how much the GP can deviate from its mean. This measurement partially solves the issue of including the off-diagonal entries in an interpretable manner since the hyperparameter controls the amplitude of both the diagonal and off-diagonal elements of the matrix. However, it is unclear how to compare two covariance matrix amplitudes with two different correlation length values. Most importantly, this measurement of uncertainty does not directly relate to the nodes of the GP themselves, only to their allowed values. In summary, there is not a singular way of quantifying the uncertainty of a GP, especially using one single number.

For the reasons discussed above in this work we will use a combination of two metrics to report the constraints on the GP. At the most basic level, we will study $\mu(z)$ itself and our constraints on its full redshift dependence. We will pay particular attention to $\mu(z=0)$ since it gives us information on the strength of the fifth force today and can easily be related to other, laboratory or astronomical constraints \cite{Ferreira:2019xrr}. In a more abstract level, we will also look at the constraints on the hyperparameter $\eta$ that describes the amplitude of the covariance matrix of the GP (i.e. the allowed deviation of the nodes from their mean).

\section{Observables and Data sets}
\label{Sect: Observables and data sets}

As previously stated, the quality of our data is just as important as our assumptions on $H(z)$ to determine our ability to constrain $\mu(z)$. In this section, we will discuss the data used in this work, as well as how we forecast what future data will be capable of. 

Let us begin by discussing the currently available data. We employ the same ensemble of data used in \citet{Ruiz_Zapatero_21}, as well as additional measurements of $f\!\sigma_8$. These can be seen in Fig. \ref{fig:data} and in the summary table in App. \ref{App: Tables}. The observables and data sets we consider are: 
\begin{figure*} 
\centering
    \includegraphics[width=\linewidth]{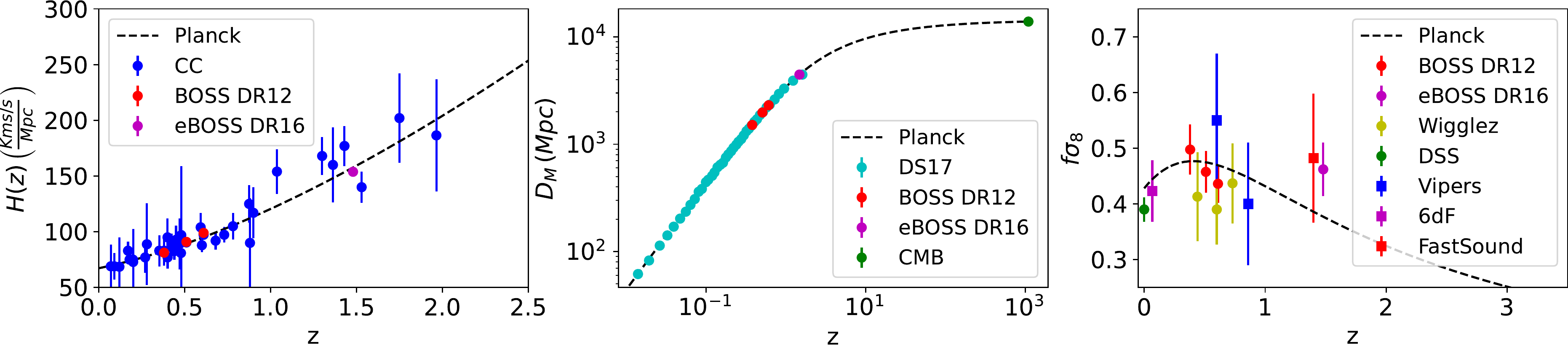} 
    \caption{Shows the data points from the different surveys used in this work across redshift for the three cosmological functions of interest $H(z)$, $D_M$ and $f \! \sigma_8$.}
   \label{fig:data}
\end{figure*}

\noindent
{\it Cosmic Chronometers} (CCs) are tracers of $dt/dz$ where $t$ is cosmic time. Since $H(z) \equiv \Dot{a}/a = -(dt/dz)/(1+z)$, a measurement of $dt/dz$ directly yields the expansion rate \citep{0106145}. Here, we use  the $H(z)$ measurements from CCs summarized in Table 1 of \citet{2011.11645}.

\noindent
{\it Type Ia supernovae} (SNe Ia) are explosions of white dwarfs \citep{Hoyle60, Colgate69}, which can be used as standard candles \citep{9907052, 0012376}.
SNe Ia obey the relationship
 $   m(z) = 5 \log_{10} D_L(z) + 25 + M \,, $
 where $m(z)$ is known as the distance modulus and $M$ is the absolute (apparent) magnitude of the SNe Ia. Knowing $M$, one can use SNe Ia to reconstruct $D_L(z)$. Here we use a compressed version of the Pantheon sample, known as "DS17", composed of 40 measurements of the distance modulus in the range $0.15\leq z\leq1.615$ \citep{1710.00845}.  We marginalize over the absolute magnitude of the supernovae as opposed to fixing its value \citep{H0_ofer}; this is equivalent to fitting the expansion rate, $E(z)= H(z)/H_0$.

\noindent
{\it Baryon acoustic oscillations} (BAOs) are set by the size of the sound horizon at the end of the drag epoch ($z \sim 1020$), \citep{Peebles70,0110414,0803.0547}
 $ r_{\rm{s}}(z)= \int^{\infty}_z [{c_{\rm{s}}/{H(z')]  \, {\rm d}z'}}$,
where $c_{\rm s}$ denotes the speed of sound. The BAO feature can be measured in the directions parallel and perpendicular to the line of sight to determine $H(z)$ and $D_M(z)$ respectively. Here we use the twelfth data release of the Baryon Oscillation Spectroscopic Survey (BOSS DR12) which forms part of the Sloan Digital Sky Survey (SDSS) III.  
In addition to this, we employ the sixteenth data release of the extended Baryon Oscillation Spectroscopic Survey (eBOSS DR16 \citealp{1703.00052}), which forms part of the Sloan Digital Sky Survey (SDSS) IV  \citep{1508.04473}. 
Finally, we make use of the \textit{Planck} 2018 measurement of the BAO angular scale at  $z_* \sim 1100$. We use the \textit{Planck} measurement from the temperature, polarization and lensing maps combined with BAO measurements denoted as TTTEEE+LowE+Lensing+BAO.

\noindent
{\it Redshift space distortions} (RSDs) are modifications to the observed redshift of a given object caused by its radial peculiar velocity \citep{kaiser87}. These leave a characteristic anisotropic imprint in the correlation function of galaxies that can be used to measure the growth of structure. Here, we use the three measurements of  $f\sigma_{\rm{8}}(z)$ from the BOSS DR12 data \citep{1607.03155}, and one value from the BOSS DR16 quasar sample.  We include full covariance matrix between the BAO and RSD measurements from these data sets \citep{1607.03155,2007.08998}. We also include the $f \! \sigma_8$ measurements reported by the WiggleZ Dark Energy Survey \citep{Drinkwater10}. Despite not being RSDs based, we also include the $f \! \sigma_8 (z=0)$ derived from the measured peculiar velocities of the Democratic Samples of Supernovae \citep{DSS}. In addition to these, we consider three additional RSD based $f \! \sigma_8$ measurements not included in \citet{Ruiz_Zapatero_21}. Namely, the $f \! \sigma_8$ measurements from the VIMOS Public Extragalactic Redshift Survey (VIPERS), the 6dF Galaxy Survey and the Subaru FMOS galaxy redshift survey (FastSound).

\noindent
Finally, we are interested in how future surveys will allow us to improve on current measurements. In order to do so, we generate synthetic data based on the forecast errors for The Dark Energy Spectroscopic Instrument (DESI).  DESI is currently taking data from the Mayall 4 meter telescope at Kitt Peak National Observatory to construct a galaxy and quasar redshift survey. We use the \citet{Font-Ribera14} forecast errors for the observables -- $H(z)$, $D_A(z)$, and $f \! \sigma_8$ -- over 18 redshift bins from 0.15 to 1.85. Then, we use the fiducial values of these quantities for the best-fit \textit{Planck} 2018 TTTEEE+LowE+Lensing+BAO $\Lambda$CDM cosmology ($\Omega^{\rm{P18}}_{\rm{M}} = 0.315$, $\Omega^{\rm{P18}}_{\Lambda} = 0.685$,   $\Omega^{\rm{P18}}_{\rm{b}}  = 0.049$, $H^{\rm{P18}}_0 = 67.36$ and $\sigma^{\rm{P18}}_{\rm{8}} = 0.811$) to generate a synthetic data set. In the following  sections, we will use this synthetic data to forecast how well a stage IV survey will do in constraining $\mu(z)$ relative to existing data.

\section{Model-dependent constraints}
\label{Sect: model-dependent}

Having discussed our modelling of $\mu(z)$ and the data we will use to constrain it, we are now at a position to start obtaining constraints for $\mu(z)$. In this section we will focus on constraints which assume a particular model for the background expansion rate $H(z)$, while modelling $\mu(z)$ as a GP. This is motivated by the results of \citet{Baker14} where it was shown that the equation of state for the energy component responsible for the accelerated expansion of the Universe would be degenerate with $\mu(z)$.

We start by considering a fiducial expansion rate -- the expansion rate given by the {\sl Planck} 18 \citep{Planck} \lcdm TTTEEE+LowE+Lensing+BAO posteriors. In this set up, we only make use of our $f\!\sigma_8$ measurements to constrain our model since we are already using {\sl Planck} 2018's posterior as a constraint on the expansion history. The parameters varied in this set up with their respective priors can be found in the first column of Tab. \ref{tab:priors}. This will give us a best case scenario and will allow us to identify a benchmark against which all other constraints can be compared. 

We then relax this assumption, removing the {\sl Planck} prior and freeing up the \lcdm parameters where,
\begin{equation} \label{eq:H_LCDM}
    H(z) =  H_0 \sqrt{\Omega_M (1+z)^3 + \Omega_R (1+z)^4 + \Omega_\Lambda} \, ,
\end{equation}
and $\Omega_M$, $\Omega_R$ and $\Omega_\Lambda$ are the cosmological matter, radiation and dark energy densities, respectively, today. We then use the measurements of $H(z)$, $D_M(z)$ and $f\sigma_8$ to constrain these parameters at the same time as we constrain $\mu(z)$. The details of this model can be found in the second column of Tab. \ref{tab:priors}.

In the next study case, to further loosen our assumptions, we chose a background rate of expansion using a general model of dark energy with an equation of state $w(a)=w_0+w_a(1-a)$ ($w$CDM). In such model the expression for the expansion rate becomes
\begin{equation} \label{eq:H_w0}
    H(z) =  H_0 \sqrt{\Omega_M (1+z)^3 + \Omega_R (1+z)^4 + \Omega_\Lambda (1+z)^{\nu(z)}}  , 
\end{equation}
where 
\begin{equation} \label{eq:nu_w0}
    \nu(z) = \frac{3(1 + w_0 + z (1+w_0+w_a))}{1+z} \, . 
\end{equation}

Similarly to the \lcdm, we consider two cases. In the first case, we use a fiducial \wcdm expansion rate -- given by Planck's \wcdm TTTEEE+lowE+lensing+BAO+SNe posteriors, where we only use $f \! \sigma_8$ measurements to constrain our model. We include SNe measurements since the TTTEEE+lowE+lensing+BAO combination considered so far is not able to place tight constraints on the equation of state on its own. The details for this model can be found in the third column of Tab. \ref{tab:priors}. In the second case, we free the expansion rate parameters (including $w_0$ and $w_a$) and use our whole suite of measurements to inform our constraints. The details of this model can found in the fourth column of Tab. \ref{tab:priors}. 

\begin{figure} 
    \includegraphics[width=1.05\linewidth]{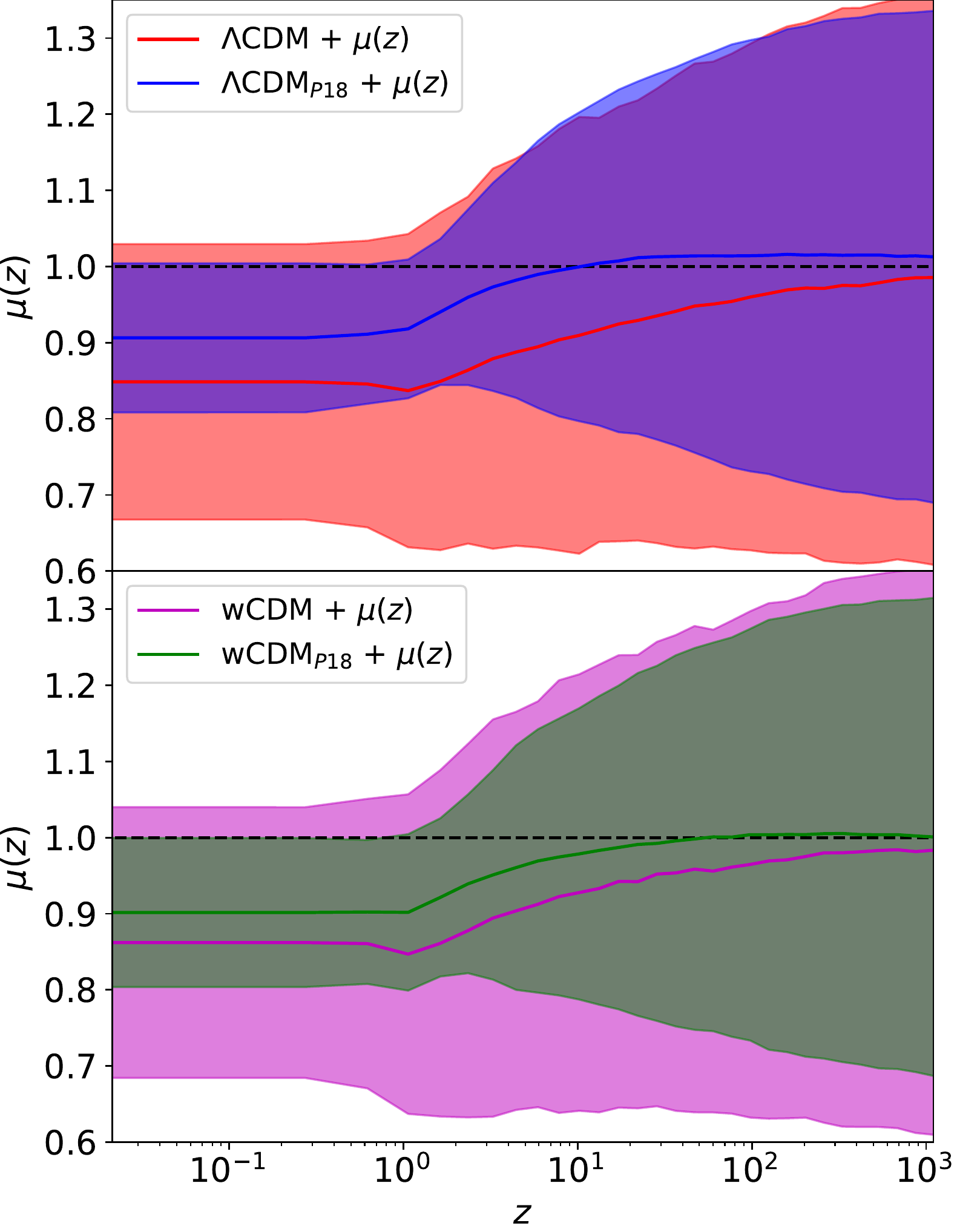} 
    \caption{Shows the obtained model-dependent constraints on $\mu(z)$. Top panel shows the constraints obtained assuming a \lcdm model for $H(z)$ both when when using {\sl Planck} 2018's $\Lambda$CDM posterior as a prior and when using current late time data to inform it (blue and red respectively). Bottom panel shows the equivalent \wcdm constraints (green and purple respectively) }
   \label{fig:Phase_1_latent}
\end{figure}

We find that regardless the model assumptions made (\lcdm or \wcdm), $\mu(z)$ is in excellent statistical agreement with the GR value $\mu(z)=1$ at all redshifts up to $1\sigma$. We find the same consistency with GR when using the {\sl Planck} 18 prior on the cosmological parameters (including $w_0$ and $w_a$ in the \wcdm case) and when freeing them. Fig. \ref{fig:Phase_1_latent} shows the constraints obtained on $\mu(z)$ in both cases, with the constraints obtained assuming \lcdm on shown in the top panel  and those assuming \wcdm in the bottom panel. In both panels we compare the contours obtained using the {\sl Planck} 2018 posterior as a prior in combination with our $f\!\sigma_8$ measurements, and by using our whole suite of measurements to inform our constraints. We can see that imposing the {\sl Planck} 2018 prior significantly reduces the uncertainty on $\mu(z)$ at all redshifts. More quantitatively (see Tab.~\ref{tab:model_depedent_constraints}), the uncertainty on $\mu_0$ decreases by roughly $\sim 35\%$ for both a \lcdm or \wcdm cosmology and, remarkably, the uncertainty in $\mu_0$ remains unchanged when using the more complex \wcdm background model. Thus, we can conclude that the combination of cosmic chronometers, BAO and SNe data are sufficiently precise to pin down the equation of state for the purpose of constraining $\mu_0$.  

\begin{figure} 
    \includegraphics[width=1.0\linewidth]{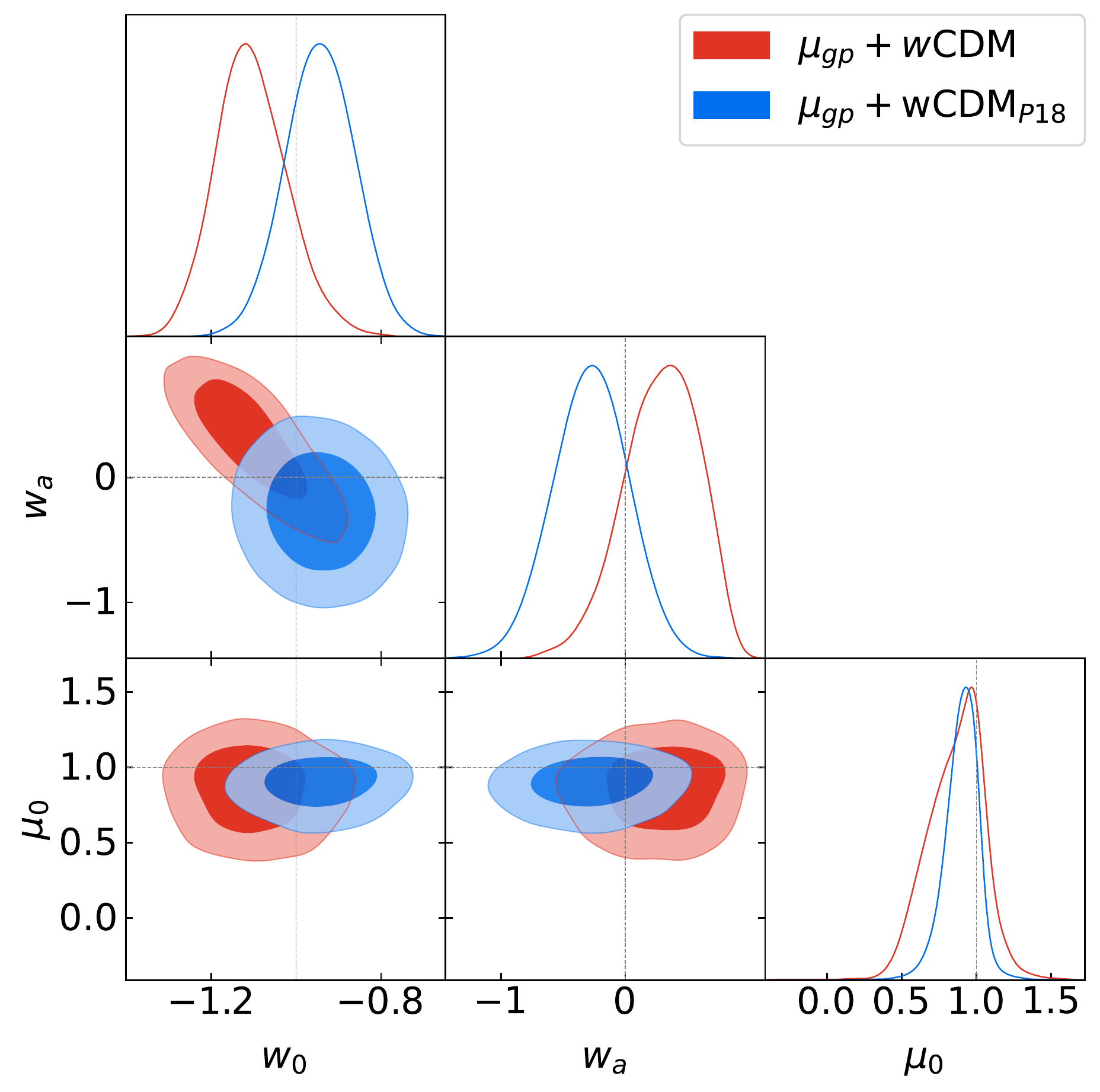} 
    \caption{Shows constraints for the cosmological parameters $w_0$, $w_a$ and $\mu_0$. Diagonal panels show 1D distributions. Off-diagonal panels show 2D distributions. In each panel we superpose the contours obtained when assuming {\sl Planck} 2018's \wcdm posterior as a prior (blue) and when marginalizing over a \wcdm background (red) given current late time data. }
   \label{fig:w0wa_triangle}
\end{figure}

It is interesting to understand this result in light of the discussion in \citet{Baker14}. There, it was shown that, while a measurement of $f\sigma_8$ at one redshift would lead to a severe degeneracy between $\mu$ and $w$, measurements at multiple redshifts combined with distance measurements could, in principle, break this degeneracy and decorrelate constraints between the two parameters. In Fig.  \ref{fig:w0wa_triangle} we see this idea in action. In this figure we show the 1D and 2D distributions for the parameters $w_0$, $w_a$ and $\mu_0$. We superpose the contours obtained when using the {\sl Planck} 2018 prior (blue) and when only using current data to constrain the \wcdm parameters (red). As we can observe, the current data contours show a degeneracy between $w_0$-$w_a$ which is not present when using the {\sl Planck} 2018 prior. However, neither $w_0$ nor $w_a$ are degenerated with $\mu_0$ in any case. 

\begin{table} 
\centering
\caption{Model-dependent constraints on $\Omega_M$, $\sigma_8$ and $\mu_0$, reporting the mean value and the $1\sigma$ errors. }
\begin{tabular}{|p{2.6cm}|p{1.9cm}|p{1.9cm}|p{1.9cm}|}
 \hline
 &  $\Omega_M$ & $\sigma_8$ & $ \mu_0$ \\
 \hline
\lcdm & $0.302 \pm 0.007$ & $0.789 \pm 0.027$ & -\\
\wcdm & $0.292 \pm 0.013$ & $0.801 \pm 0.034$ & -\\
$\mu(z)+\Lambda$CDM$_{P18}$ & $0.314 \pm 0.007$ & $0.811 \pm 0.006$ & $0.904 \pm 0.123$ \\
$\mu(z)+w$CDM$_{P18}$ & $0.306 \pm 0.008$ & $0.821 \pm 0.014$ & $0.899 \pm 0.123$ \\
$\mu(z)+\Lambda$CDM & $0.302 \pm 0.007$ & $0.878 \pm 0.127$ & $0.850 \pm 0.191$ \\
$\mu(z)+w$CDM & $0.29 \pm 0.016$ & $0.887 \pm 0.127$ & $0.862 \pm 0.190$ \\

\hline
\end{tabular}
\label{tab:model_depedent_constraints}
\end{table}

We further note that the uncertainty in $\mu(z)$ increases as we look at higher redshifts but not excessively so. Two factors are at play here. First, since the data are a non-local function of $\mu(z)$ (i.e. $\mu(z)$ needs to be integrated to solve for $f$ in Eqn. \ref{eq:growth}), they allow us to place constraints on higher redshift values of $\mu(z)$. In addition to this, we are marginalizing over the hyperparameters of the Gaussian process. This means that the data at lower redshifts can put a constraint on the amplitude and correlation length of the GP's kernel. This effectively limits the variance of the GP even in regions with no data.

\begin{figure} 
    \includegraphics[width=1.0\linewidth]{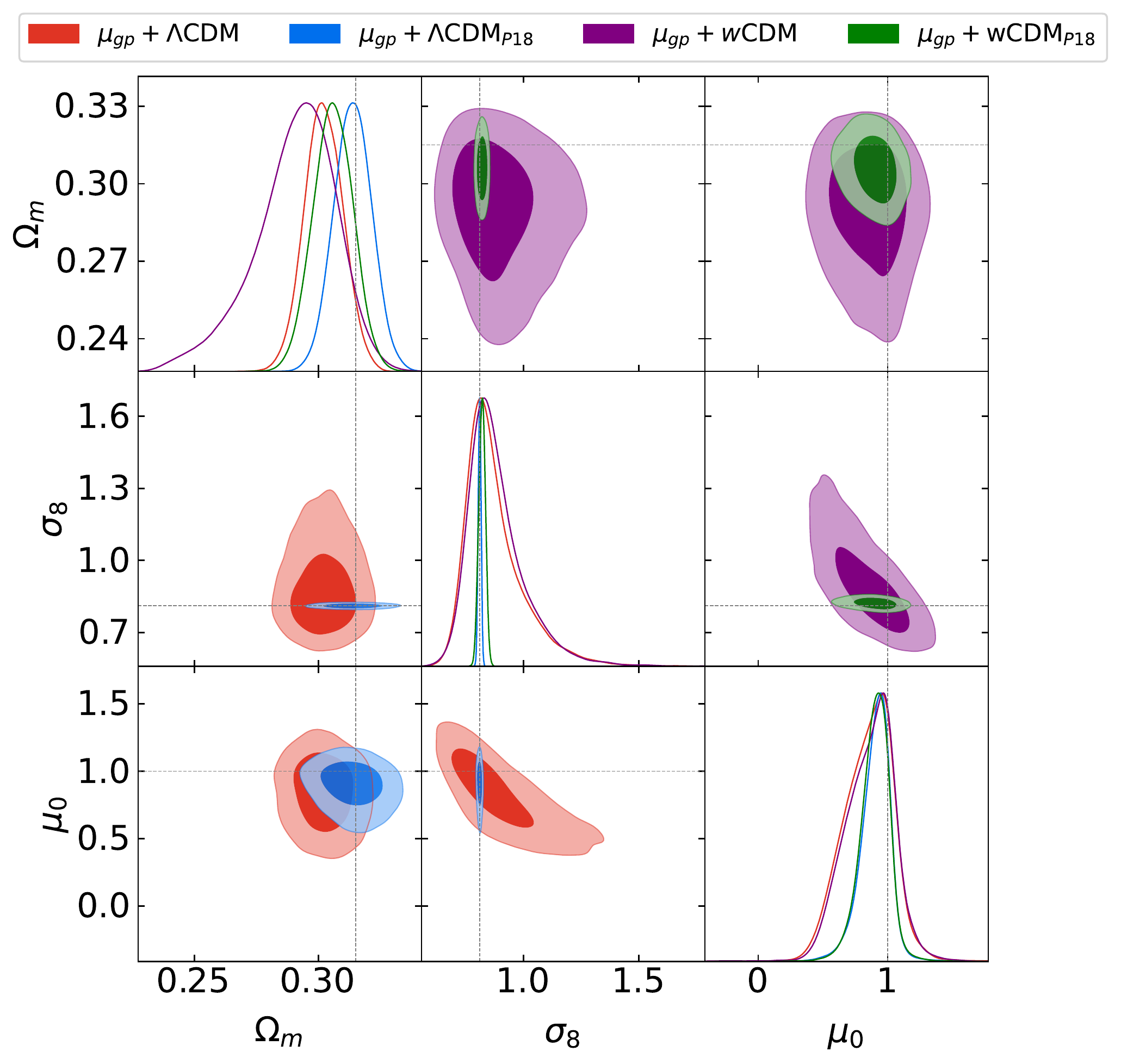} 
    \caption{Shows constraints for the cosmological parameters $\Omega_M$, $\sigma_8$ and $\mu_0$. Diagonal panels show 1D distributions. Off-diagonal panels show 2D distributions. Bottom triangle shows the constraints obtained when assuming a \lcdm background both when imposing the {\sl Planck} 2018's $\Lambda$CDM posterior as a prior (blue) and when using current late time data to inform it (red). The top triangle shows the equivalent constraints when a \wcdm background was assumed instead (green and purple respectively).}
   \label{fig:small_triangles}
\end{figure}

We have seen that assuming a \wcdm for $H(z)$ as opposed to $\Lambda$CDM model does not degrade our constraints on $\mu(z)$. It is then interesting to explore the relationship between $\mu(z)$ and other cosmological parameters of our models, particularly $\Omega_M$ and $\sigma_8$. Fig. \ref{fig:small_triangles} shows the 1D and 2D contours for the parameters $\Omega_M$, $\sigma_8$ and $\mu_0$ obtained when assuming the \lcdm and \wcdm models to parameterize $H(z)$. In each panel we superpose the results obtained when assuming {\sl Planck} 2018's posterior as a prior for the expansion rate as opposed to letting background data inform the constraints. We show the associated numerical constraints in Tab. \ref{tab:model_depedent_constraints}. We also display the constraints obtained by fitting a $\Lambda$CDM and $w$CDM model while keeping $\mu(z) = 1$ (i.e. GR) for context.

 Looking at Eq. \ref{eq:growth} one would expect a great degeneracy between $\Omega_M$ and $\mu(z)$. However, if we look at the bottom left corner panel ($\Lambda$CDM) and top right panel ($w$CDM) of Fig. \ref{fig:small_triangles} we can see how information about the background breaks this degeneracy. Therefore, it is not clear that a better constraint on one will lead to an improvement on the other. 
 
\begin{figure} 
    \includegraphics[width=\linewidth]{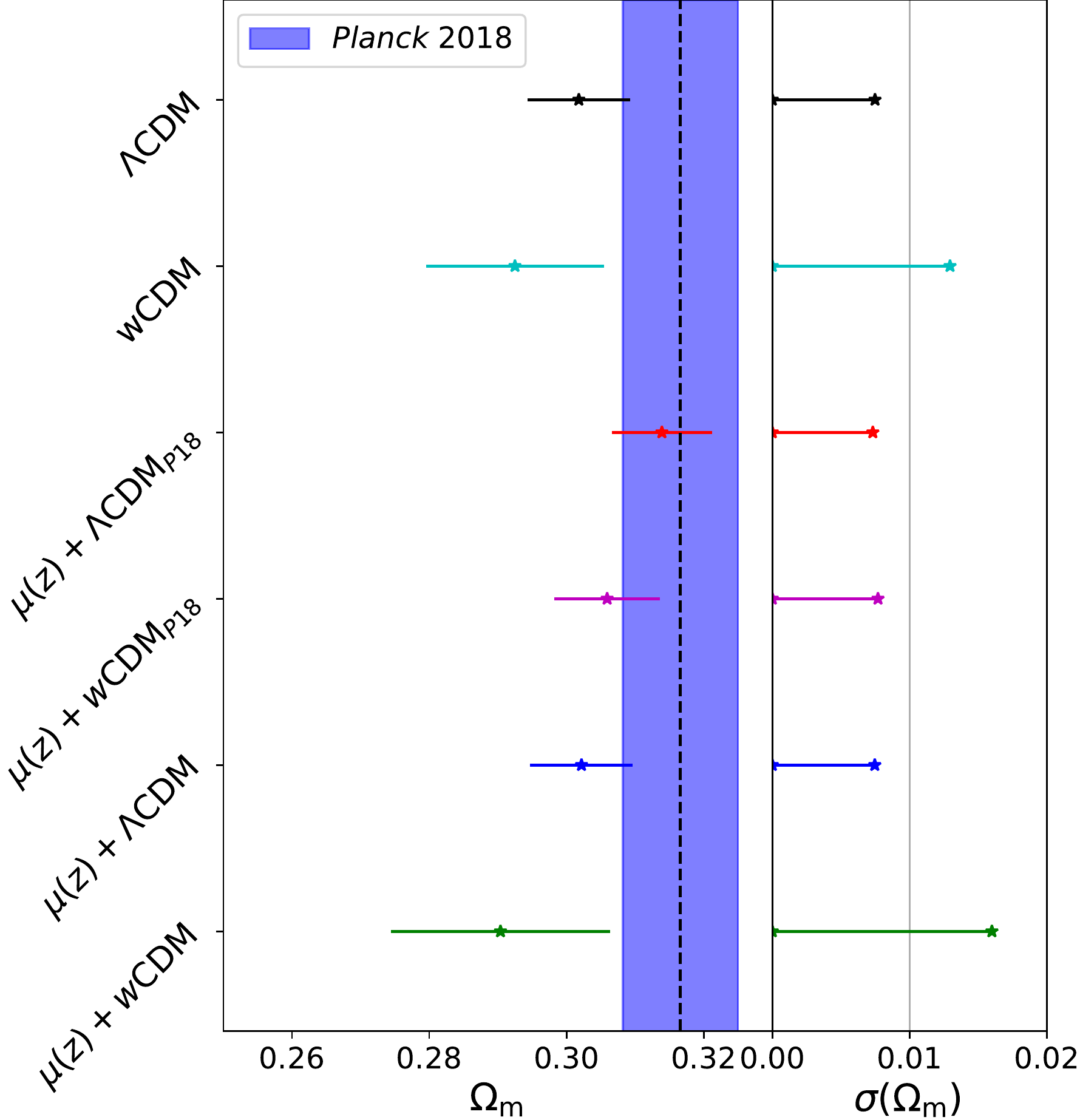} 
    \caption{Shows the constraints obtained for $\Omega_M$ for each model considered in this work. Side panel shows the uncertainty of each constraint. }
   \label{fig:Omega_ms}
\end{figure}

 We show our constraints on $\Omega_M$ for the different models in Fig. \ref{fig:Omega_ms}, including constraints for the $\Lambda$CDM and \wcdm models when keeping $\mu(z) = 1$ (i.e. GR) for reference. Regardless of whether we assume a $\Lambda$CDM or $w$CDM model for $H(z)$ we obtain a slightly lower value for $\Omega_M$ than the one obtained by {\sl Planck} 2018 (and the one obtained using {\sl Planck} 2018's posterior as a prior). Nonetheless, once the size of the error bars is taken into account, the constraints are in reasonable statistical agreement (less than 1.5 $\sigma$ tension). Moreover, assuming $w$CDM systematically results in a lower yet statistically compatible constraint of $\Omega_M$ than assuming a $\Lambda$CDM model. Finally, it is interesting to note that introducing $\mu(z)$ barely degrades the constraint on $\Omega_M$ if the background is \lcdm. On the contrary, for a \wcdm model, introducing $\mu(z)$ leads to a $\sim 20 \%$ larger error bar on $\Omega_M$. This is caused by the fact that freeing the equation of state reduces the ability of the background to constrain $\Omega_M$ and, thus, the $w$CDM $\Omega_M$ constraint increasingly depend on the growth data to inform its value. 
 
Moving to $\sigma_8$, current growth data cannot break the degeneracy between $\sigma_8$ and $\mu_0$. This can be seen in the middle panel of the bottom row and the right panel of the middle row. Therefore, when assuming a model for $H(z)$, the bottleneck in constraining $\mu_0$ is how well we know $\sigma_8$. This explains why our constraints on $\mu(z)$ drastically improve when imposing the {\sl Planck} 2018 prior since it imposes a much tighter constraint on $\sigma_8$, breaking the degeneracy with $\mu(z)$.

\section{Model-independent constraints}
\label{Sect: Double GP}

We now proceed to further relax our assumptions about the background expansion rate by promoting $H(z)$ to a GP. We do so by following the methodology developed in \citet{Ruiz_Zapatero_21}. More specifically, we model $H(z)$ as
\begin{equation} \label{eq:H_gp}
    H(z) = A_0 {H}^{\rm{P18}}(z) (1+\delta H_{gp}) \, , 
\end{equation}
where $A_0$ is a free parameter, $H^{\rm{P18}}(z)$ is the Hubble rate for our $\Lambda$CDM {\sl Planck} 18 best-fit fiducial cosmology (see Sect. \ref{Sect: a Gaussian process for muz}), and $\delta H_{gp}$ is a relative deviation that we model as a Gaussian Process. This is a Bayesian approach to GP's in which one marginalizes simultaneously over the GP itself and its mean. This type of approach has recently been discussed in\citet{How_to_use_GP}.

This means that our inference process now involves two GP's. This allows us to measure the degeneracy between modifications of the expansion history and the Poisson equation in the prediction of $f \sigma_8$ without having to assume a particular model.

However, becoming fully model-independent comes at the cost of no-longer being able to constrain $\Omega_M$ with measurements of background quantities. This is because $H(z)$ is no longer a function of cosmological parameters. Thus, we have no independent way of constraining $\Omega_M$ apart from the relationship between $H(z)$ and $f\sigma_8(z)$. Revisiting Eq. \ref{eq:growth}, we can also see that we are now faced with an unbreakable degeneracy between $\Omega_M$ and $\mu(z)$. In order to deal with this degeneracy, in this section we consider the new, combined parameter
 \begin{eqnarray}
 {\tilde \mu}(z)=\frac{\Omega_M}{ \Omega^{\rm{P18}}_M} \mu(z) \, , 
 \end{eqnarray}
 where $\Omega^{\rm{P18}}_M$ is the {\sl Planck} 18 TTTEEE+LowE+Lensing+BAO, \lcdm best fit value of $\Omega_M$.

In order to solve Eqns. \ref{eq:growth} and \ref{eq:distance} when considering two GPs, we employ the same combination of  numerical methods as in \citet{Ruiz_Zapatero_21} (where we also modelled $H(z)$ as a GP), albeit with some modification. In \citet{Ruiz_Zapatero_21} we assigned a node of the GP to each node of the numerical grid used to solve the growth equation and the comoving distance integral. This approach becomes very computationally expensive when we introduce a second GP. In order to make our model more computationally efficient, we decouple the number of nodes in the numerical integration schemes from the number of nodes used for each GP, linearly interpolating where necessary. This allows us to significantly reduce the number of parameters of the model while preserving the necessary numerical accuracy. Reducing the number of nodes in the GP's means that the degeneracy between the remaining nodes is reduced. This latter aspect is particularly helpful when using HMC which is most efficient when the parameters are as uncorrelated as possible. The end result of reducing the number of parameters and the degeneracy between them is a substantial speed-up in the time needed for the sampler to converge. 

\begin{figure}
    \includegraphics[width=1.0\linewidth]{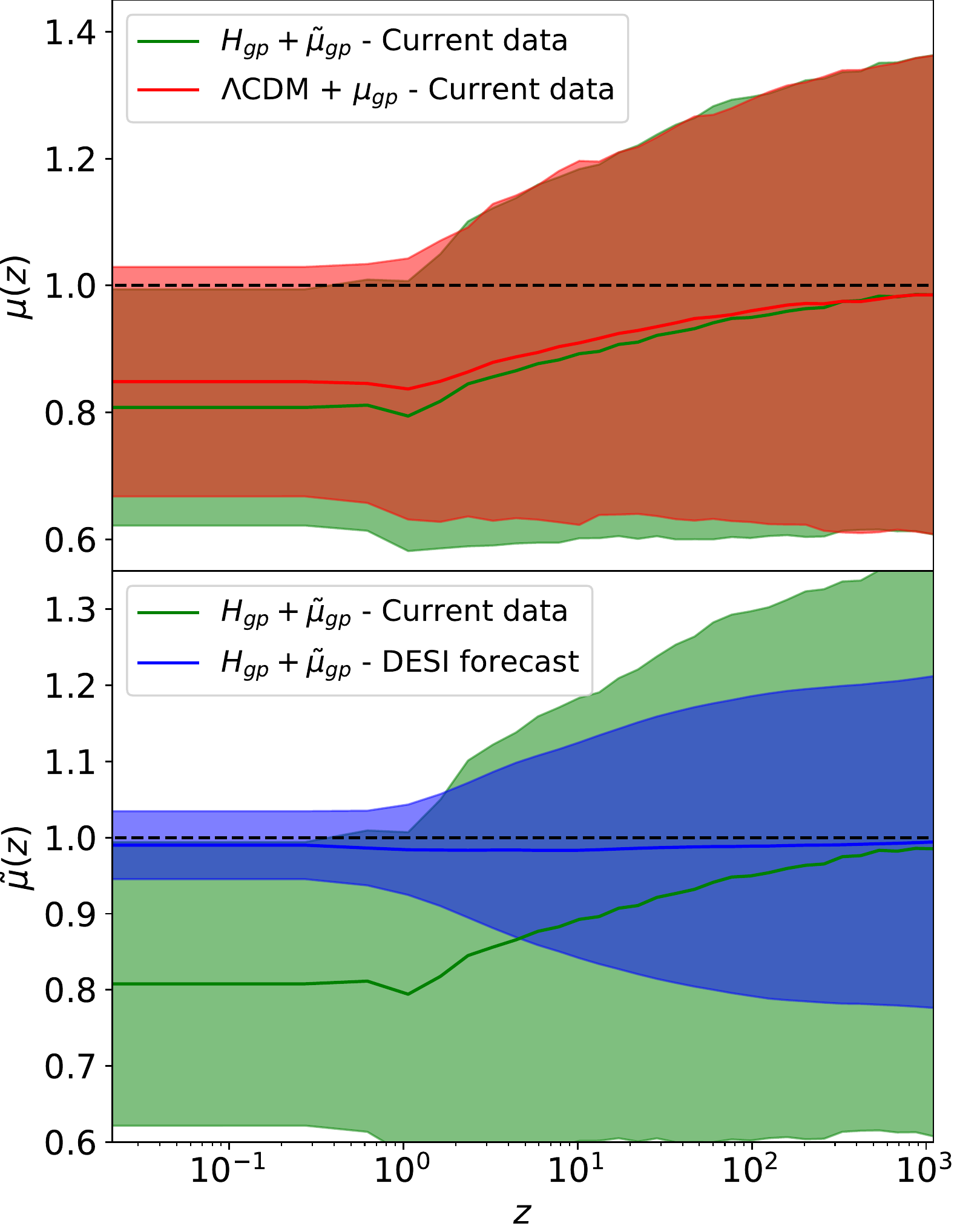} 
    \caption{Top panel: shows constraints on $\mu(z)$ for current data when assuming the $\Lambda$CDM model to model background expansion of the Universe (red) and when using a second GP (green). Note that when using a second GP the quantity being constrained is $\tilde{\mu}(z)$ as opposed to $\mu(z)$. 
    Bottom panel: shows the constraints obtained on $\tilde{\mu}(z)$ when using a second GP to model $H(z)$ for both current data (green) and mock DESI data (blue).}
   \label{fig:Phase2_latent}
\end{figure}

\begin{figure}
    \includegraphics[width=1.0\linewidth]{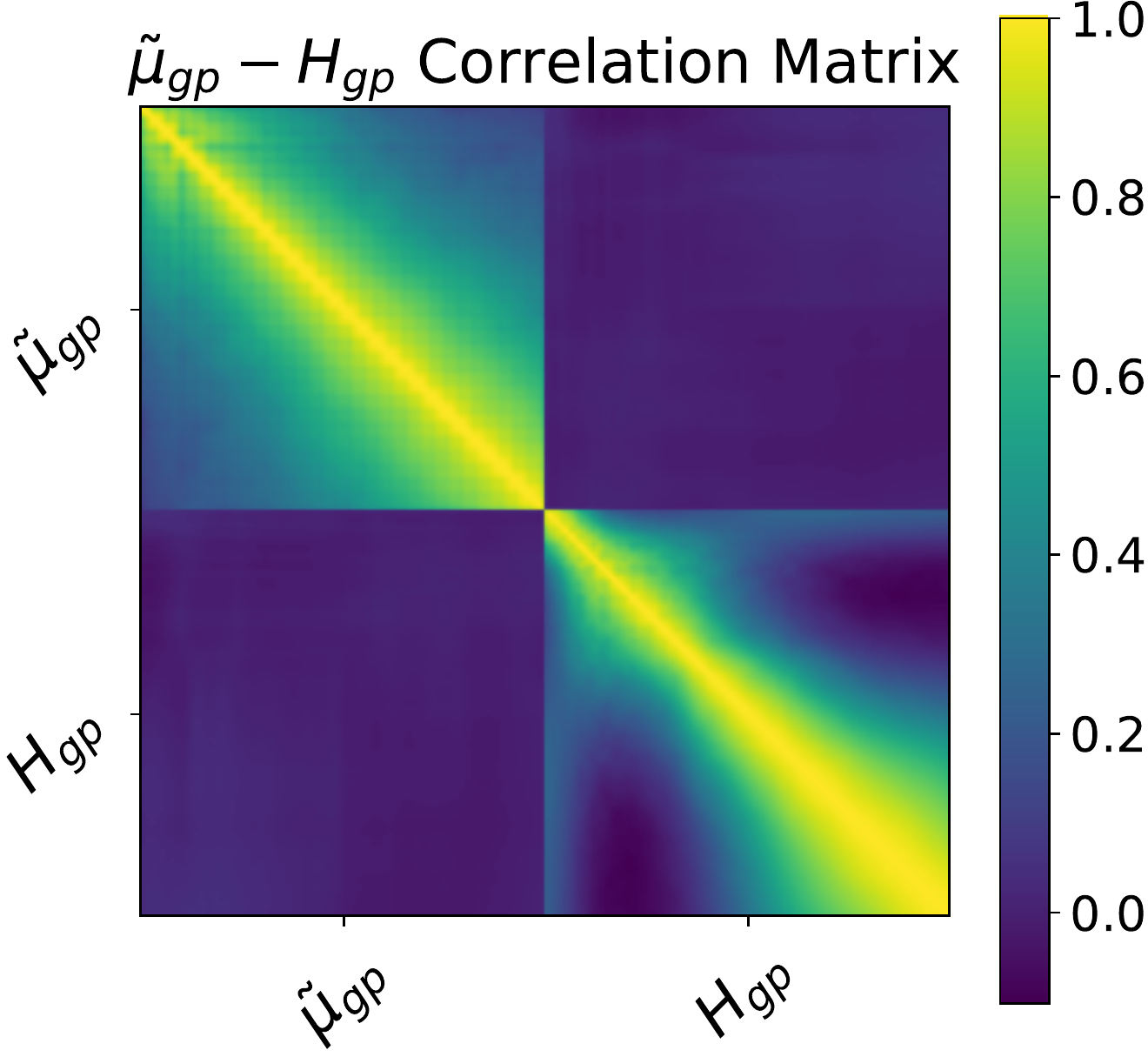} 
    \caption{Shows correlation coefficients between the nodes of the GP on $H(z)$ and the GP on $\mu(z)$. This plot can be seen as a generalization of Fig.~\ref{fig:w0wa_triangle}, showing that the expansion rate and the modifications of the linear growth are already independent with the constrain level of current data.}
   \label{fig:corr_plot}
\end{figure}

We show the obtained model-independent constraints for $\tilde \mu(z)$ in Fig. \ref{fig:Phase2_latent}. On the one hand, in the top panel of the figure, we can observe that the model-independent constraints on $\tilde \mu(z)$ are only marginally worse than the model-dependent constraints on $\mu(z)$ ($5\% - 10\%$ depending on whether we consider the $\Lambda$CDM or \wcdm model). This means that, even when completely relaxing our assumptions about $H(z)$, current data have enough constraining power to break the degeneracy between $H(z)$ and $\tilde{\mu}(z)$. This can be further seen in the correlation matrix between the GP's nodes of $\tilde\mu(z)$ and $H(z)$. Fig. \ref{fig:corr_plot} shows that, although $\mu(z)$ and $H(z)$ nodes have a great degree of auto-correlation (as expected for a GP), the correlation coefficients between both quantities are never larger than $5\%$. This can be seen as a generalization of the lack of correlation we observed between the background parameters and $\mu_0$  in Sect. \ref{Sect: model-dependent}. Moreover, we can see that $H(z)$'s low redshift nodes are much less correlated with high redshift nodes than those of $\tilde{\mu(z)}$.

It is important to bear in mind that these are constraints on ${\tilde \mu}(z)$, not on $\mu(z)$. Converting constraints on  ${\tilde \mu}(z)$ into constraints on $\mu(z)$ requires a measurement of $\Omega_M$. However, in the process of freeing $H(z)$ we have lost all of our knowledge of $\Omega_M$. Thus, an external, model-independent measurement of $\Omega_M$ would be needed to transform ${\tilde \mu}(z)$ constraints into $\mu(z)$ constraints. The constraints on ${\tilde \mu}(z)$ should therefore be understood as the most optimistic model-independent constraint on $\mu(z)$ possible given current data; i.e. the case for which we have a perfect model-independent measurement of $\Omega_M$. 

Finally, we find $\sigma_8 = 0.886 \pm 0.138$ when a second GP is used to model $H(z)$. This means that not assuming a \lcdm or \wcdm model for the expansion history degrades our $\sigma_8$ constraint by around $\sim 10\%$. Nonetheless, the degree of correlation between $\sigma_8$ and $\tilde{\mu}_0$ remains virtually identical to that of model-dependent analyses. Thus, model-independent constraints on $\tilde{\mu}(z)$ will also benefit greatly from ways of tightening their constraint on $\sigma_8$, just as we saw in the model-dependent case. We will discuss this further in the next section when considering our analysis of mock DESI data.

\section{Discussion} 
\label{Sect: Conclusions}

\begin{figure}
    \includegraphics[width=1.0\linewidth]{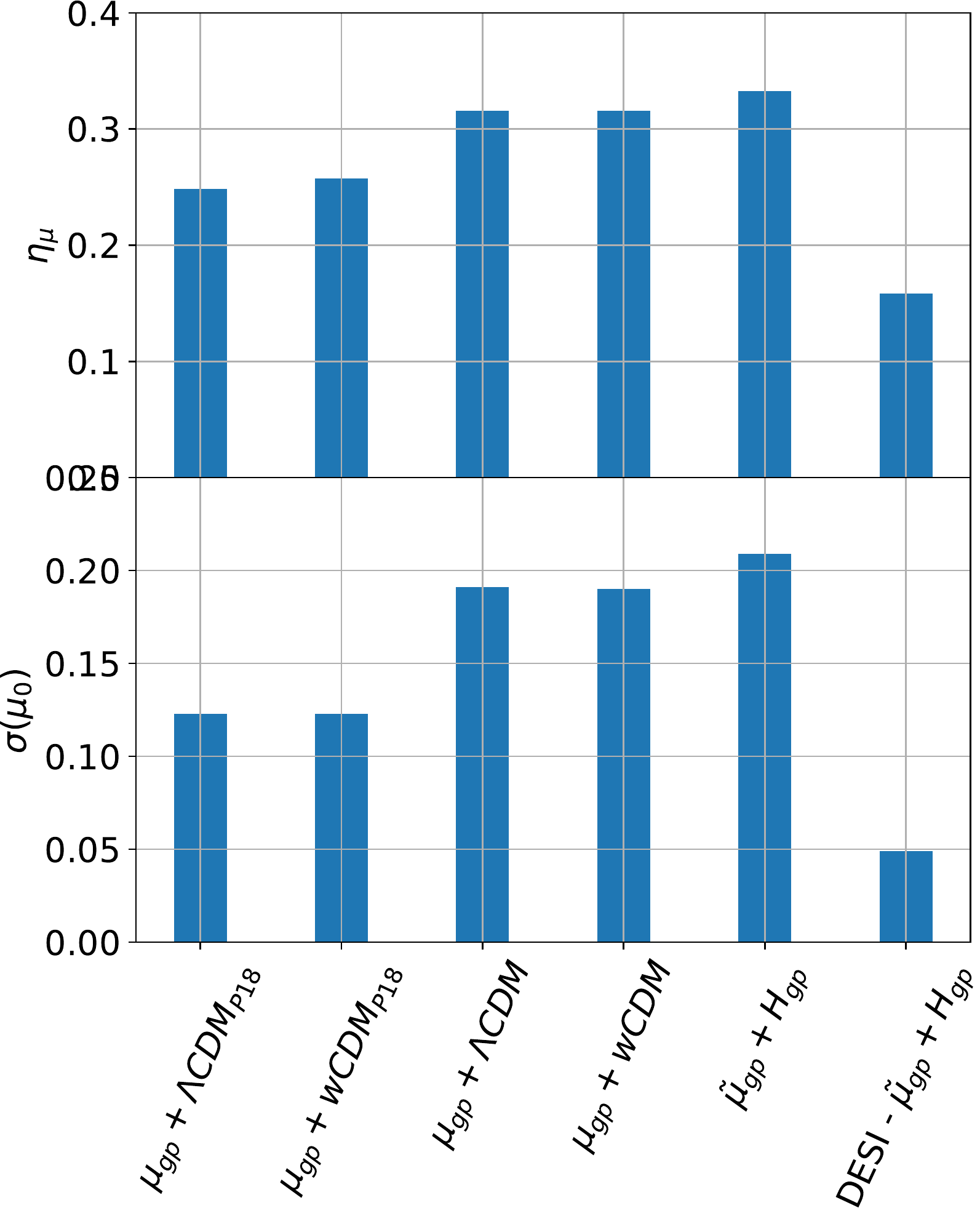} 
    \caption{Shows a comparison of different measures of uncertainty in $\mu(z)$ between the different models considered in this work. The top panel shows the mean value of the amplitude of the covariance matrix of $\mu(z)$ for each model. The bottom panel shows the uncertainty in $\mu(z=0) \equiv \mu_0$. Note that when a second GP is considered to model $H(z)$ (i.e. last two entries)  $\tilde{\mu}(z)$ is shown as opposed to $\mu(z)$.}
   \label{fig:bar_plot}
\end{figure}

In this paper we have assessed the importance of our current knowledge of the expansion rate history on our ability to constrain $\mu(z)$ in a model-independent manner. As was argued in
\citet{Simpson:2009zj}  and \citet{Baker14}, the assumptions that go into the modelling the Hubble rate as a function of redshift, $H(z)$ will impact constraints on $\mu(z)$ from the growth rate of structure. It was shown that the more conservative (or looser) the model for $H(z)$, the weaker the constraints on $\mu(z)$ should be.

We have found that, however, current constraints on the expansion rate from cosmic chronometers, supernovae and BAO data are sufficiently tight that the assumptions made about the underlying background model are not important when constraining $\mu(z)$. To show this, we have used a completely general form for $\mu(z)$ (a Gaussian Process), and quantified whether assuming a simple equation of state for Dark Energy ($w=-1$), or a more general equation of state of the form $w=w_0+w_a(1-a)$ affects the final constraints on $\mu(z)$. We also considered a completely general form for $H(z)$ which we also modelled as a Gaussian Process. In this case, we are faced with a fundamental degeneracy between $\mu(z)$ and $\Omega_M$ and thus, we present our results in terms of ${\tilde \mu}(z)=\Omega_M\mu(z)/\Omega^{P18}_{M}$ where we recall that $\Omega^{P18}_{M}$ is the best fit value of $\Omega_M$ for the {\sl Planck} 2018 TTTEEE+LowE+Lensing+BAO analysis of the $\Lambda$CDM model.

As discussed in Section \ref{Sect: Interpreting a Gaussian Process}, we summarize our results on the constraints on $\mu(z)$ using two statistics. On the one hand, we look at the uncertainty in $\mu_0 \equiv \mu(z=0)$ as it directly relates to the strength of any possible fifth force today. On the other hand, we consider the mean value of the amplitude of the Gaussian process covariance matrix, $\eta_\mu$, which is an abstract measurement of the uncertainty of the Gaussian process through its whole domain. 

We present the corresponding results in Fig. \ref{fig:bar_plot}. Reassuringly, we find that the two statistics offer us the same picture: the less assumptions we make on the expansion history, the more uncertainty on $\mu(z)$. However, it is extremely important to stress that the loss in constraining power is marginal. Comparing assuming a \lcdm vs \wcdm model, we find that it makes effectively no difference and there is no degradation in our constraints on $\mu(z)$. Even when a second GP is used to model $H(z)$ the constraint is only a few percentages larger.

Focusing on $\mu_0$, we find that  $\sigma(\mu_0)\simeq 0.12$ if we assume {\sl Planck} 2018's posterior as a prior, for either the $\Lambda$CDM or $w$CDM model. This uncertainty increases to $\sigma(\mu_0)\simeq 0.19$ if instead of imposing {\sl Planck} 2018's posterior as a prior we use our collection of late time $H(z)$, $D_M(z)$ and $f\sigma_8$ measurements to inform our constraints. The main difference between assuming {\sl Planck} 2018 posteriors and using late-time data to inform our models is that the former provides us with a much tighter constraint on $\sigma_8$, the main bottle-neck when constraining $\mu(z)$ in a model-dependent fashion. Looking at the model-independent constraint, we find that $\sigma({\tilde \mu}(z)) \simeq 0.21$. 

If we instead look at the constraints on $\eta_\mu$, we find the exact same trend as in $\mu_0$. While one would expect the two statistics to agree, $\mu_0$ only probes the GP at $z=0$ while $\eta_\mu$ contains information about the whole GP domain. We find that for our best-case scenario, in which we assume {\sl Planck} 2018's \lcdm background, $\eta_\mu = 0.25$. Letting late time data inform a \lcdm model instead returns $\eta_\mu = 0.32$. Furthermore, if we assume a \wcdm model, we find  $\eta_\mu = 0.26$ when using the {\sl Planck}'s posteriors to pin it and $\eta_\mu = 0.32$ when letting late time data inform it. Finally, we find $\eta_\mu = 0.33$ in the model-independent case.  

The fact that constraints on $\mu$ are (relatively) insensitive to our parametrization of $H(z)$ is not unexpected. This is because current background data is powerful enough to constrain $H(z)$ independently of the assumptions made. In the analysis of \citet{Ruiz_Zapatero_21}, we found that constraints on $\Omega_M$ from the growth rate were not strongly dependent on our modelling choices of the Gaussian process on $H(z)$.

There have been other attempts at constraining $\mu(z)$. In \citet{Planck} an uncertainty of $\sigma(\mu_0)\simeq 0.25$ was found under the assumption that $\mu$ evolves as  $\mu(z) - 1 \propto [1-\Omega_M(z)]$. However, different assumptions about the specific time dependence of $\mu$ (e.g. $\mu(z)\propto a^n$) lead to constraints that are strongly dependent on the choice of $n$\cite{Mueller:2016kpu}, with $\mu_0$ in the range $\sigma(\mu_0)\in(0.04, 1.5)$. Assuming that the modified Poisson equations arises from scalar-tensor theories, one can use the tools of Effective Field Theory \cite{Raveri:2021dbu} or simply assume specific classes of models \cite{Traykova:2021hbr} to obtain $\sigma(\mu_0)\simeq 0.25$. As we can see, our methodology returns stronger constraints with $\sigma(\mu_0)\in(0.12, 0.19)$ depending on the strength of the assumptions made on $H(z)$. We note however that it can be misleading to directly compare $\sigma(\mu_0)$ as they can be heavily dependent on the underlying model and choice of data sets one is using.

It is instructive to see how much our constraints will improve with future data. As an example, we choose the specifications for the DESI data set, described in Sect. \ref{Sect: Observables and data sets}, and combine it with the {\sl Planck} 2018 CMB BAO measurement to pin down the GP on $H(z)$ at high redshift . Our analysis of DESI mock data shows that we will obtain constraints on $\tilde \mu(z)$ (i.e. with a model-independent $H(z)$) which are twice as tight as with current data when assuming either a $\Lambda$CDM or $w$CDM background. This is in spite of the DESI constraint on $\sigma_8$ being about six times wider than {\sl Planck} 2018's. The reason behind this improvement in constraining power boils down to the fact that DESI alone will offer nearly twice as many measurements on $f \! \sigma_8$ as the number considered in this work over a larger redshift window. Moreover, DESI  $f \! \sigma_8$ measurement will have significantly smaller errors bars than currently available ones. It is particular important to focus on the smaller size of said error bars relative to the expected dynamic range of $f \! \sigma_8$ in the redshift window probed. This will greatly help break the degeneracy between the amplitude of   $f \! \sigma_8$ (given by $\sigma_8$) and its shape (given by $\mu(z)$ in the presence of background data to pin down $\Omega_M$) present in current data.

Finally, there are several avenues through which the results and methodology presented here could be further explored. One can ask the question: how well do we need to measure $\Omega_M$ to obtain a competitive model-independent constraint on $\mu(z)$ with current data. Using  propagation of errors; $\sigma(\mu) / \mu = \sqrt{(\sigma(\tilde{\mu}) / \tilde{\mu})^2 + (\sigma( \Omega_M) / \Omega_M)^2}$, we find that model-independent measurement of $\Omega_M$ to $10 \%$ precision would be enough to match model-independent constraints on $\mu(z)$ to model-dependent constraints with current data. Similarly, a percentage model-independent measurement of $\Omega_M$ would allow us to constrain $\mu(z)$ to virtually the same precision as $\tilde \mu(z)$. This measurement of $\Omega_M$ would need to be independent from the model assumed for the background expansion and for the parametrization of the Poisson equation. Future works could attempt to obtain an alternative model-independent constraint on $\Omega_M$ to break the $\tilde{\mu}(z)-\Omega_M$ degeneracy found in this methodology.

\section*{Acknowledgements}
\textit{Author contributions}: All authors contributed to the development and writing of
this paper.

We would like to thank Dan Foreman-Mackey and Andreu Font-Ribera for their helpful comments. We acknowledge support from the Beecroft Trust. CGG and PGF acknowledge funding from the European Research Council (ERC) under the European Unions Horizon 2020 research and innovation programme (grant agreement No 693024). DA acknowledges support from the Science and Technology Facilities Council through an Ernest Rutherford Fellowship, grant reference ST/P004474. JRZ is supported by an STFC doctoral studentship.

The analysis made use of the software tools \textsc{PyMC3} \cite{Pymc3}, \textsc{Theano} \cite{Theano},  \textsc{NumPy} \citep{np}, \textsc{Matplotlib} \citep{plt}, \textsc{CLASS} \citep{Class}, GetDist \citep{getDist}.

\appendix
\section{Tables} \label{App: Tables}

In this appendix we display the large tables that would have interrupted the reading flow of the main body paper. Table \ref{tab:data} contains the data sets used in this work.

\begin{table*} 
\centering
\caption{Data sets used in our analysis, listing the probe, the redshift range of the probe, the choice of observable and and the size of the data vector. }
\begin{tabular}{|p{3.5cm}|p{2cm}|p{2.2cm}|p{0.8cm} p{.8cm} p{0.8cm}|p{1cm} |  }
 \hline
Data set &  Probe & Redshifts &\multicolumn{3}{|c|}{Observable} & Data Points\\
&  & & $H(z)$ & $D_m(z)$ & $f \! \sigma_8 $ & \\
 \hline
 CC's \citep{2011.11645} &  Cosmic Chronometers & 0.07 - 1.965 & $\checkmark$ & $\times$ & $\times$ & 33\\
 
Pantheon DS17\citep{1710.00845} & SNe Ia & 0.38 - 0.61 & $\times$ & $\checkmark$ & $\times$ & 40\\
 
 BOSS DR12 \citep{1607.03155} & BAO+RSD & 0.38 - 0.61 & $\checkmark$ & $\checkmark$ & $\checkmark$ & $3\times3$\\

 eBOSS DR16 \citep{2007.08991} & BAO+RSD & 1.48 & $\checkmark$ & $\checkmark$ & $\checkmark$  & $1\times3$ \\ 
 
 Wigglez \citep{1204.3674} & RSD & 0.44 - 0.73 & $\times$ & $\times$ & $\checkmark$  & 3 \\
 
 Vipers \citep{Vipers} & RSD & 0.60 - 0.86 & $\times$ & $\times$ & $\checkmark$  & 2 \\
 
 6dF \citep{6dF} & RSD & 0.067 & $\times$ & $\times$ & $\checkmark$  & 1 \\
  
 FastSound \citep{FastSound} & RSD & 1.4 & $\times$ & $\times$ & $\checkmark$  & 1 \\
 
 DSS \citep{2105.05185} & RSD & 0 & $\times$ & $\times$ & $\checkmark$  & 1 \\
 
 \textit{Planck} 2018  \citep{Planck} & CMB & 1090.30 & $\times$ & $\checkmark$ & $\times$  & 1 \\
 \hline
 DESI \citep{Font-Ribera14} & BAO+RSD & 0.15 - 1.85 & $\checkmark$ & $\checkmark$ & $\checkmark$ & $3\times18$\\
 \hline

\end{tabular}
\label{tab:data}
\end{table*}

Table \ref{tab:priors} contains the prior distributions assumed for each for our models. In general, the priors are chosen broad enough to prevent biasing our results. In particular, the priors on the hyperparameters of the GP on $\mu(z)$ ($\eta_\mu \and l_\mu$), common in all the studied cases. As discussed in \citet{Ruiz_Zapatero_21}, when using gradient based methods it is best-practice to use smooth priors unless there's physical limit on the values that the parameter can take (e.g. $\Omega_M \in [0, 1]$).

Thus, the prior of the amplitude of the GP, $\eta_\mu$,  is a half normal distribution $\mathcal{N}_{1/2}(0, 0.5)$; i.e. centered at $0$ with $0.5$ standard deviation. On the other hand, the correlation length $l_\mu$ has an uniform prior $U(0.01, 6)$. The reason for a uniform prior (i.e. not smooth) is two fold. On the one hand, when sampling $l_\mu$ it is extremely important to avoid small values in order to avoid volume effects (See Eqn .\ref{eq:QuadExp}). On the other hand, we do not want to down/up-weight a particular correlation scale for the nodes of GP. 

Moving on to the cosmological parameters, only $\Omega_M$ has an uniform prior $U(0, 1)$ to enforce the physical limits on the values of the parameter. All the others have normal distributions whose details can be found in Table~\ref{tab:priors}. For the cases with a {\sl Planck} 2018 prior, we use the values quoted in \citet{Planck}. In particular, for the $\Lambda$CDM$_{P18} + \mu_{GP}$ case (second column), we used the  TTTEEE+lowE+lensing+BAO \lcdm constraints (last column of Tab. 2 in \citet{Planck}), while for the \wcdm$_{P18} + \mu_{GP}$ case (forth column), we used the  TTTEEE+lowE+lensing+BAO+SNe \wcdm constraints (first column of Tab. 6 in \citet{Planck}). Note that in the \wcdm case the constraints also include SNe data which is not present in the \lcdm constraints. This is because TTTEEE+lowE+lensing+BAO data cannot constrain \wcdm models by itself. Note that for both the \lcdm and the \wcdm models we fix $\Omega_R = 9.245 \times 10^{-5}$. $\Omega_\Lambda$ is then derived using $ 1 = \Omega_M+\Omega_R+\Omega_\Lambda$. 

We must also consider a number of nuisance parameters needed to model the specific data sets chosen for this work. For instance, in order to relate the luminosity curves of the Pantheon data set to luminosity distances one needs to know the value of the absolute magnitude of the supernovae, $M$. In this work we choose the agnostic way and marginalize over $M$, assuming a normal prior $\mathcal{N}(-19.2, 1)$, which encompasses both \citet{riess_42}'s and \citet{Planck}'s $H_0$ values. On the other hand, we make extensive use of measurements of both parallel and perpendicular BAO measurements. In order to relate these measurements to $H(z)$ and $D_M(z)$ one needs to know the value of the sound horizon at either drag ($r_d$) or recombination ($r^*$) epochs. In order to  obtain $r_d$ and $r^*$ we use a modified version of the Eisenstein and Hu fitting formula \citep{EisHu_rd, Nesseris_rd} given by
\begin{eqnarray}  \label{eq:Nesseris_rd}
    r_d \approx \frac{45.5337 \ln{(7.20376/\omega_m)}}{\sqrt{1+9.98592 \omega_b^{0.801347}}} \, \rm{Mpc}, 
\end{eqnarray}
where $\omega_m = \Omega_M (H_0/100)^2$ and $\omega_b = \Omega_b (H_0/100)^2$. Then, noting that the ratio between  $r_d$ and $r^*$ can be approximated as a function exclusively of $\Omega_b$, we derive the fitting formula 
\begin{eqnarray}  \label{eq:rd_rs}
    \left(\frac{r_d}{r^*}\right) (\Omega_b)   \approx 1.11346 - 2.7985 \Omega_b + 16.5111 \Omega_b^2 \, .
\end{eqnarray}

Hence, combining Eqns. \ref{eq:Nesseris_rd} and \ref{eq:rd_rs} we can obtain a prediction for $r^\star$. This approach is capable of reproducing the CLASS \lcdm predictions for  $r_d$ and $r^*$ to an average of $1.5\%$ precision within the considered $\Omega_M \in [0.1, 0.6]$ and $\Omega_b \in [0.03, 0.07]$. Since the \wcdm model we consider doesn't include early dark energy, we can also use Eqns. \ref{eq:Nesseris_rd} and \ref{eq:rd_rs} to predict the values of $r_d$ and $r^*$ in such case.
On the other hand, when a second GP is used to model $H(z)$,  $\Omega_M$ is absorbed into the GP on $\mu(z)$ to form $\tilde{\mu}(z)$. This disallows us from following the same approach to obtain $r_d$ and $r^*$ as when assuming a \lcdm or \wcdm model. In this scenario we sample $r_d$ directly as a parameter from $\mathcal{N}(145, 5)$. Then, to get $r^*$ we use Eqn. \ref{eq:rd_rs} as a function of $r_d$ and $\Omega_b$ using the same  $\Omega_b$ as in the \lcdm and \wcdm case.

\begin{table*} 
    \caption{Priors used for the different parameters of the models considered in this work. The first column shows the complete list of parameters. $U$ stands for a uniform distribution; $\mathcal{N}(a, b)$ and $\mathcal{N}_{1/2}(a, b)$,  for a normal and half-normal distribution, respectively, centered at $a$ and with standard deviation $b$.  Empty entries represent parameters not sampled by the model.} 
    \centering
    \begin{tabular}{|p{1cm}|p{2.5cm}|p{2.5cm}|p{2.5cm}|p{2.5cm}|p{2.5cm}|}
    \hline
     & $H_{\Lambda \rm{CDM} , \rm{P18}}+\mu_{gp}$ & $H_{\Lambda \rm{CDM}}+\mu_{gp}$ & $H_{\rm{wCDM}, \rm{P18}}+\mu_{gp}$ & $H_{\rm{wCDM}}+\mu_{gp}$ &  $H_{gp}+\tilde{\mu}_{gp}$  \\
    \hline
    $A_0$ & - & - & - & - & $\mathcal{N}(1.0, 0.2)$  \\
    $\eta_{\rm{H}}$   &  - & - & - & - & $\mathcal{N}_{1/2}(0, 0.2)$\\
    $l_{\rm{H}}$   &  - & - & - & - & $U(0.01, 6)$ \\
    $\eta_{\rm{\mu}}$ & $\mathcal{N}_{1/2}(0, 0.5)$ & $\mathcal{N}_{1/2}(0, 0.5)$ & $\mathcal{N}_{1/2}(0, 0.5)$ & $\mathcal{N}_{1/2}(0, 0.5)$ & $\mathcal{N}_{1/2}(0, 0.5)$ \\
    $l_{\rm{\mu}}$ & $U(0.01, 6)$ & $U(0.01, 6)$ & $U(0.01, 6)$ & $U(0.01, 6)$ & $U(0.01, 6)$ \\
    \hline
    $\Omega_{\rm{m}}$ & $\mathcal{N}(0.316, 0.008)$ & $U(0,1)$ & $\mathcal{N}(0.307, 0.011)$ & $U(0,1)$  & - \\
    $\Omega_{\rm{b}}$ & - & $U(0.03, 0.07)$ & - & $U(0.03, 0.07)$  & $U(0.03, 0.07)$ \\
    $H_0$ & $\mathcal{N}(67.27, 0.6)$ & $\mathcal{N}(70, 5)$ &  $\mathcal{N}(68.31, 0.82)$ & $\mathcal{N}(70, 5)$ & -\\
    $\sigma_8$ & $\mathcal{N}(0.811, 0.007)$ & $\mathcal{N}(0.8, 0.5)$  & $\mathcal{N}(0.82, 0.011)$  &  $\mathcal{N}(0.8, 0.5)$ & $\mathcal{N}(0.8, 0.5)$ \\
    $w_0$ & - & - & $\mathcal{N}(-1, 0.5)$ & $\mathcal{N}(-0.957, 0.08)$  & -\\
    $w_a$ & - & - & $\mathcal{N}(0, 0.5)$ & $\mathcal{N}(-0.29, 0.3)$  & -\\
    \hline
    $M$ & - & $\mathcal{N}(-19.2, 1)$ & - & $\mathcal{N}(-19.2, 1)$ & $\mathcal{N}(-19.2, 1)$   \\
    $r_d$ & - & Derived & - &  Derived & $\mathcal{N}(150, 5)$ \\
    $r^*$ & - & Derived & - &  Derived & Derived \\
    \hline
\end{tabular}
\centering
\label{tab:priors}
\end{table*}

\bibliography{biblio.bib}
\end{document}